\documentclass[twocolumn]{aastex631}

\usepackage[normalem]{ulem}

\graphicspath{{./}{figures/}}

\def\Halpha{\mbox{H\hspace{0.1ex}$\alpha$}\,}
\def\kms{\hbox{km$\;$s$^{-1}$}}

\def\deg{\hbox{$^\circ$}}
\def\mAA{m\AA\,}
\def\FeI{\ion{Fe}{1}}
\def\FeIVISline{\ion{Fe}{1}~6569\,\AA\,}
\def\FeIIRline{\ion{Fe}{1}~8538\,\AA\,}
\def\SiIIRline{\ion{Si}{1}~8536\,\AA\,}
\def\HeIIRline{\ion{He}{1}~10830\,\AA\,}
\def\CaIR{\ion{Ca}{2}~8542\,\AA\,}
\def\CaHK{\ion{Ca}{2}~H\&K\,}
\def\logtau{$\log\tau_{\mathrm{500}}$\,}

\def\blos{$B_{\mathrm{LOS}}$\,}

\def\temp{$T$\,}
\def\vlos{$V_{\mathrm{LOS}}$\,}
\def\vturb{$V_{\mathrm{turb}}$\,}
\def\utw{\ensuremath{\smash{\rlap{\lower5pt\hbox{$\sim$}}}}}
\def\udtw{\ensuremath{\smash{\rlap{\lower6pt\hbox{$\approx$}}}}}
\def\imeanobs{$I_{\mathrm{obs}}^{\mathrm{mean}}$\,}
\def\iref{$I_{\mathrm{ref}}$\,}

\def\deltab{$\Delta \lambda_{\mathrm{B}}$\,}
\def\deltad{$\Delta \lambda_{\mathrm{D}}$\,}

\begin{document}


\title{Does \Halpha{} Stokes~$V$ profiles probe the chromospheric magnetic field? An observational perspective

\footnote{Not yet released}}

\author[0000-0001-5253-4213]{Harsh Mathur}
\affiliation{Indian Institute of Astrophysics, II Block, Koramangala, Bengaluru 560 034, India}

\author[0000-0002-0465-8032]{K. Nagaraju}
\affiliation{Indian Institute of Astrophysics, II Block, Koramangala, Bengaluru 560 034, India}

\author[0000-0003-0585-7030]{Jayant Joshi}
\affiliation{Indian Institute of Astrophysics, II Block, Koramangala, Bengaluru 560 034, India}

\author[0000-0002-4640-5658]{Jaime de la Cruz Rodríguez}
\affiliation{Institute for Solar Physics, Dept. of Astronomy, Stockholm University, AlbaNova University Centre, 10691 Stockholm, Sweden}

\begin{abstract}

We investigated the diagnostic potential of the Stokes~$V$ profile of the \Halpha{} line to probe the chromospheric line-of-sight (LOS) magnetic field (\blos{}) by comparing the \blos{} inferred from the weak field approximation (WFA) with that of inferred from the multi-line inversions of the \CaIR{}, \SiIIRline{} and \FeIIRline{} lines using the STiC inversion code.
Simultaneous spectropolarimetric observations of a pore in the \CaIR{} and \Halpha{} spectral lines obtained from the SPINOR at the Dunn Solar Telescope on the 4th of December, 2008 are used in this study.
%
%
The WFA was applied on the Stokes~$I$ and $V$ profiles of \Halpha{} line over three wavelength ranges viz.: around line core ($\Delta\lambda=\pm0.35$~\AA), line wings ($\Delta\lambda=[-1.5, -0.6]$ and $[+0.6, +1.5]$~\AA) and full spectral range of the line ($\Delta\lambda=\pm1.5$~\AA) to derive the \blos{}.
We found the maximum \blos{} strengths of $\sim+800$ and $\sim+600$~G at \logtau{} = $-$1 and $-$4.5, respectively in the pore.
The morphological map of the \blos{} inferred from the \Halpha{} line core is similar to the \blos{} map at \logtau{} = $-$4.5 inferred from multi-line inversions.
%
%
The \blos{} map inferred from the \Halpha{} line wings and full spectral range have a similar morphological structure to the \blos{} map inferred at \logtau{} = $-$1.
The \blos{} estimated from \Halpha{} using WFA is weaker by a factor of $\approx 0.53$ than that of inferred from the multi-line inversions.
%

\end{abstract}


\label{sect:intro}
\section{Introduction}
Between the bright solar surface and million degrees hot corona, the chromosphere is one of the most dynamic and complex layers of the solar atmosphere.
Understanding the magnetic coupling of the chromosphere with the photosphere, transition region and corona may reveal how mass and energy are supplied to the corona and heliosphere.
Therefore, simultaneous magnetic field measurements at multiple heights of the solar atmosphere are important.

Magnetic field measurements in the photosphere are routinely made, and significant progress has been made in the techniques that allow for chromospheric magnetic field measurements.
The most widely used lines to probe the magnetic field of the solar chromosphere are the \CaIR{} and \HeIIRline{} lines \citep[][]{2017SSRv..210...37L}.
While using these lines to probe the chromospheric magnetic field has its advantages like the relatively well-understood line formation, and they can be interpreted utilizing non-local thermodynamic equilibrium (non-LTE) inversion codes \citep[e.g.][]{2000ApJ...530..977S, 2008ApJ...683..542A, 2016ApJ...830L..30D, 2019A&A...623A..74D, 2022A&A...660A..37R}, they also have some limitations.
The \HeIIRline{} line forms in a narrow range of heights in the upper chromosphere, and line formation depends on coronal and transition region EUV radiation \citep{1997ApJ...489..375A,2016A&A...594A.104L}.
The \CaIR{} line forms from the upper photosphere to the mid-chromosphere, however, the \ion{Ca}{2} ion gets ionized to the \ion{Ca}{3} ion in flaring regions, and hence the \CaIR{} line itself may sample deeper layers of the solar atmosphere \citep[][]{Kerr_2016, 2018ApJ...860...10K, 2019A&A...631A..33B}.

On the other hand, the \Halpha{} line seems to probe a relatively wider temperature range.
\citet[][]{2002ApJ...572..626C} have shown that the \Halpha{} opacity in the upper chromosphere is determined by the ionization degree and radiation field.
Using 3-D radiative transfer calculations, \citet[][]{2012ApJ...749..136L} have shown that ionization degree and the radiation field are insensitive to local temperature variations over time but determined by mass density.
More recently, \citet{2019A&A...631A..33B} have also confirmed that the \Halpha{} line retains opacity even in flaring active regions by synthesizing spectra using a 3D rMHD simulation.
Therefore, the \Halpha{} line always retains opacity in the chromosphere.

In spite of the diagnostic capabilities of the \Halpha{} line, very few spectropolarimetric observations have been reported in the literature.
For example, \citet{1971SoPh...16..384A} estimated vertical magnetic field gradient using simultaneous measurements of the \ion{Fe}{1}~6302.5~\AA\, and \Halpha{} lines.
Simultaneous spectropolarimetric observations from the \Halpha{} line and lines of \ion{Fe}{1} atom have been reported in the literature to compare the photospheric and chromospheric magnetic fields in sunspots \citep[][]{2004ApJ...606.1233B, 2005PASJ...57..235H, 2008ApJ...678..531N}.
Radial variation of the line-of-sight magnetic field in the chromosphere and photosphere of a sunspot were discussed by \citet{2020JApA...41...10N}.
Magnetic fields in prominences were diagnosed using spectropolarimetric observations of the \Halpha{} line \citep[][]{2005ApJ...621L.145L}.
\citet{2022A&A...659A.179J} studied linear and circular polarization signals near the North and South Solar Limb and inferred the LOS magnetic field using the weak field approximation (WFA).

The reason for so few spectropolarimetric observations of the \Halpha{} line have been reported and why this has not been a preferred line for the chromospheric magnetometry is because it is challenging to model the \Halpha{} line.
Though the Stokes~$V$ signal is dominated by the Zeeman effect, both the intensity and polarization profiles of this line are sensitive to the 3D radiation field.
Furthermore, in case of weakly magnetized atmospheres, the Stokes~$Q$~\&~$U$ signals are sensitive to atomic polarization \citep{2010MmSAI..81..810S,2011ApJ...732...80S}, making this line difficult to model using the currently available inversion codes which adopt 1.5D plane-parallel geometry.
It is also important to note that in the weak field regime (when Zeeman Splitting, \deltab{}, is  much smaller than the Doppler width, \deltad),
the amplitude of circular polarization is proportional to the ratio of \deltab{} to \deltad{}, and the linear polarization is proportional to the square of that ratio \citep[for more details see page 405 of][]{2004ASSL..307.....L}.
This ratio, owing to the light weight of the Hydrogen atom-thus correspondingly large Doppler width, is typically small compared to that in the case of heavier atoms like Calcium.
%

In a study by \citet{2004ApJ...603L.129S} who calculated 1-D response functions of Stokes parameters of the \Halpha{} line and showed that it exhibits significant sensitivity to the photospheric magnetic fields in addition to the chromospheric magnetic fields.
However, it has been shown by \citet{2012ApJ...749..136L} that 1-D radiative transfer is not a good approximation to model the \Halpha{} line, treatment of radiative transfer in 3-D is necessary.
When the radiative transfer is treated in 3-D, the \Halpha{} line traces the chromospheric magnetic features like fibrils, and it has been shown that it is a good chromospheric diagnostic.
This finding is further supported by recent work by \citet{2019A&A...631A..33B}.
%

In this study, we explore the diagnostic potential of the \Halpha{} line to probe the chromospheric magnetic field using spectropolarimetric observations simultaneously recorded in the \Halpha{} and \CaIR{} lines.
We compare the magnetic field inferred using the WFA method on the Stokes~$I$ and $V$ profiles of the \Halpha{} spectral line with the stratification (from the photosphere to the chromosphere) of line-of-sight (LOS) magnetic field inferred using the inversions of the \CaIR{} line using a non-LTE inversion code.

\section{Observations} \label{sect:observations}

\begin{figure}[htbp]
\centering
            \resizebox{0.4\textwidth}{!}{\includegraphics[]{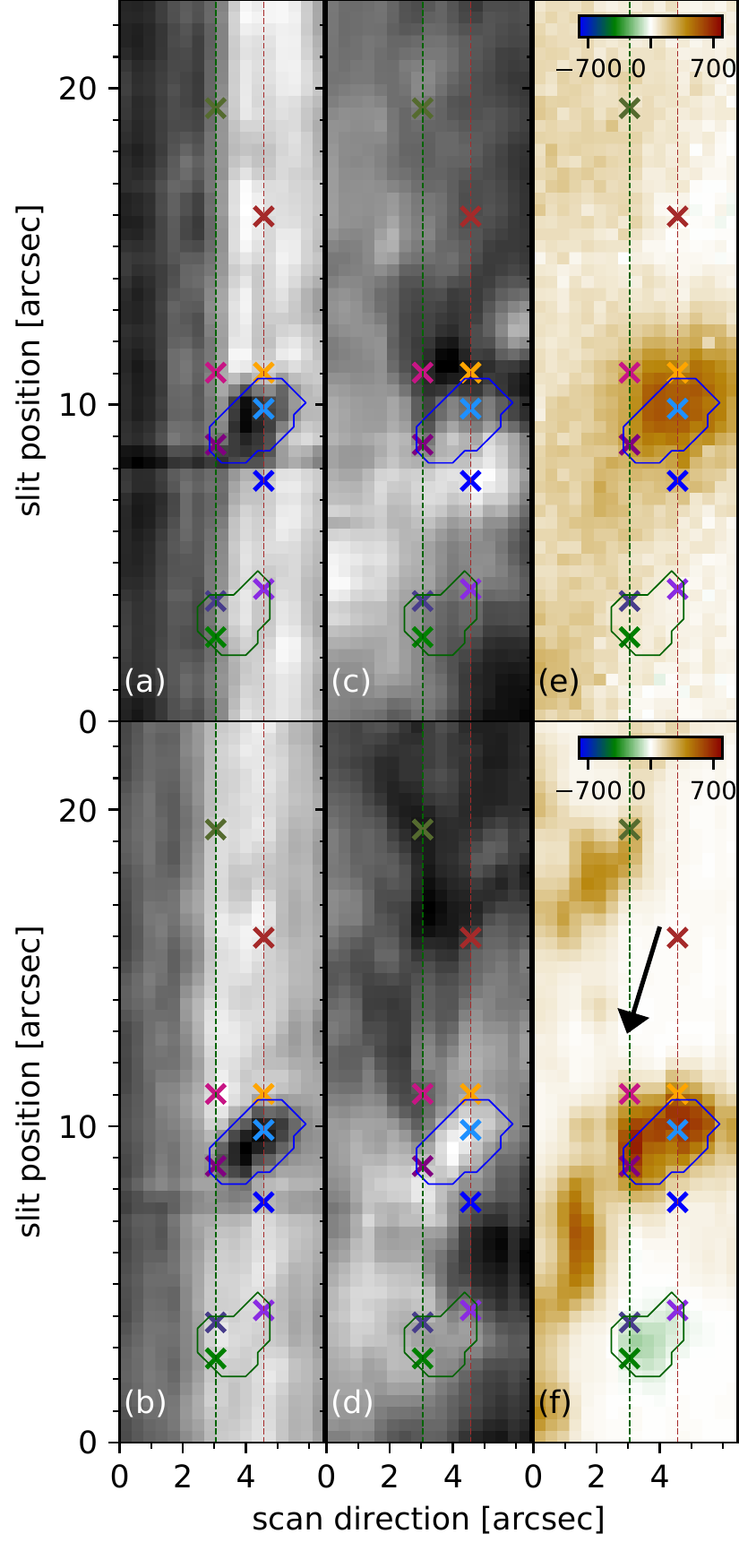}}
      \caption{
      Field of view with a pore in the center as observed by the Dunn Solar Telescope on 4th December 2008: (a) Far wing image at an offset of $-7$~\AA\, from the \CaIR{} line, (b) Far wing image at an offset of $+7$~\AA\, from the \Halpha{} line (c) \CaIR{} line core image (d) \Halpha{} line core image. Panels (e) and (f) show LOS magnetic field maps derived under weak field approximation from the \CaIR{} line within the spectral range of $\pm$0.25~\AA\, and Milne-Eddington inversions of the \FeIVISline{} line, respectively. The blue contours show the location of the pore made using intensity thresholding on the far wing image of \Halpha{} line in panel (b). The green contours show the location of the region with the negative polarity of the magnetic field seen in panel (f). The arrow indicates the disc-center direction.
             }
         \label{fig:FOV}
   \end{figure}

The observations were made with the Spectro-Polarimeter for Infrared and Optical Regions \citep[SPINOR:][]{2006SoPh..235...55S} instrument at the Dunn Solar Telescope \citep{1969S&T....38..368D} at the Sacramento Peak Observatory in the \Halpha{} and \CaIR{} lines simultaneously.
The \FeIVISline{} line was recorded in the \Halpha{} spectrum and the \SiIIRline{} and \FeIIRline{} lines were recorded in the \CaIR{} spectrum.
The spectral sampling of the \Halpha{} and \CaIR{} lines data are 22~\mAA{} and 33~\mAA{}, respectively.
%
%
The pixel scale corresponds to $\sim0\farcs38$ on the solar surface along the slit.
The observed field of view (FOV) consists of a pore centered at North 28$^o$.3 and East 16$^o$.7 in Stonyhurst Heliographic coordinates system on December 4, 2008, starting at 15:35 UT with a viewing angle $\cos\theta = \mu = 0.8$.
Here, $\theta$ is the angle between the LOS direction and the local surface normal.
Twenty spectropolarimetric raster scans of 20 slit positions with a step size of $0\farcs375$ were recorded; however, only four scans had good seeing conditions, and only the first scan is used in this study.
Adaptive optics \citep[AO;][]{2000SPIE.4007..218R} were used during the observations.
The data were corrected for dark and flat field variations and instrumental polarization.
The details are given in appendix~\ref{sect:datareduc}.
In spite of strictly simultaneous observations, there was a spatial shift, due to atmospheric refraction, between the images at these two wavelengths, which were taken care of by co-aligning the \Halpha{} data with the \CaIR{} data by cross-correlating the far wing images of the \Halpha{} and \CaIR{} lines.
The signal-to-noise ratio (SNR) in the \Halpha{} data is higher ($5\times10^{3}$) than in the \CaIR{} data ($10^{3}$).

An overview of observations is shown in Fig.~\ref{fig:FOV}.
%
%
A pore almost centrally located in the FOV can be seen in the far wing images of the \CaIR{} and \Halpha{} lines (see panels (a) and (b) of Fig.~\ref{fig:FOV}).
The pore morphologically has a different shape in the \CaIR{} and \Halpha{} lines and is brighter in the \Halpha{} line core image (see panels (c) and (d) of Fig.~\ref{fig:FOV}).
Panels (e) and (f) show the LOS magnetic field (\blos{}) inferred from the \CaIR{} and \FeIVISline{} lines, respectively.
More details on the methods to infer the \blos{} are discussed in section~\ref{sect:methods}.
%
%
%
%
There is an opposite polarity region near the pore visible in the photosphere that is absent in the WFA \blos{} map of the chromosphere.
%
%
%

\begin{figure*}[htbp]
    \centering
    \includegraphics[width=\textwidth]{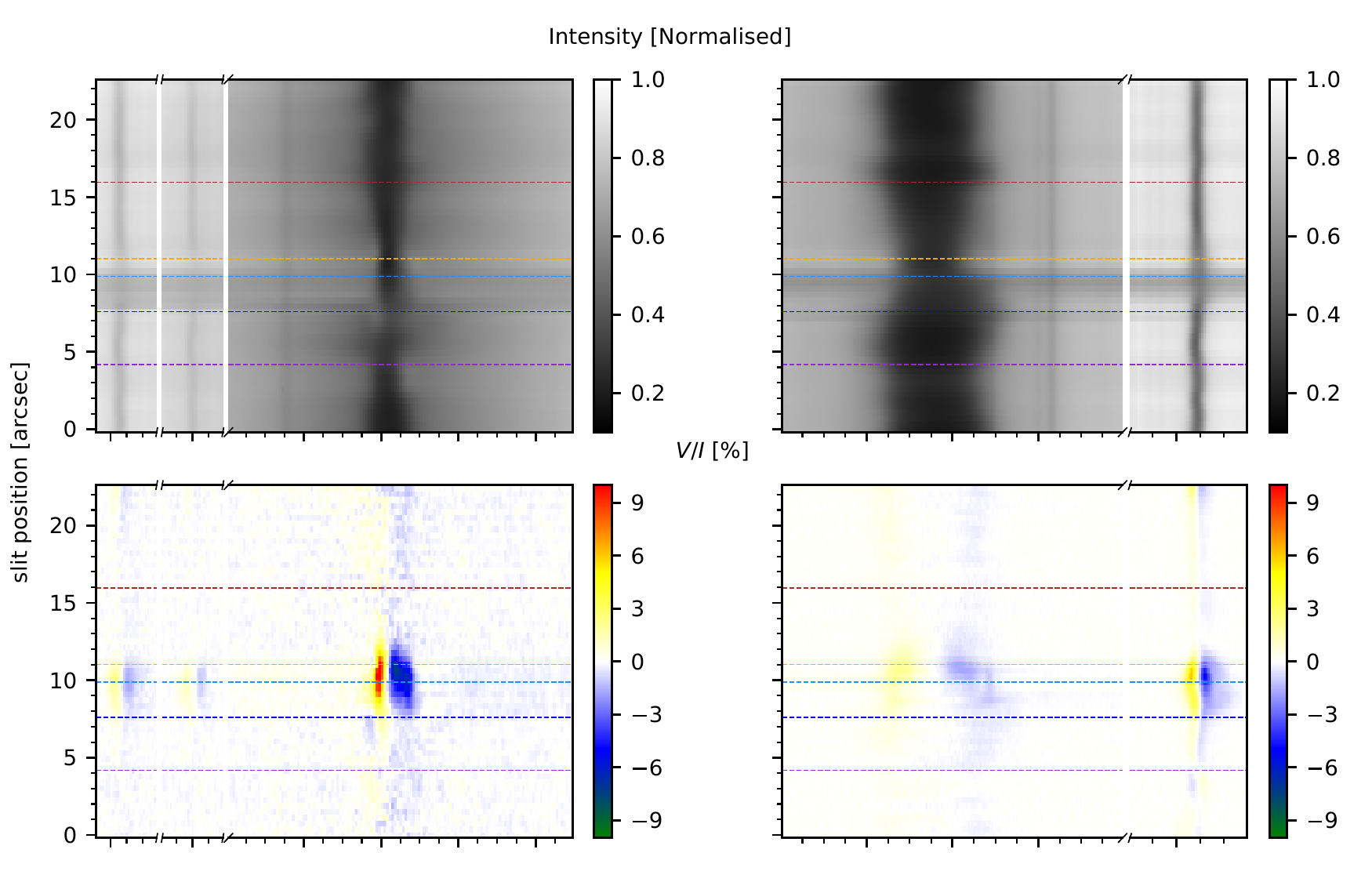}
    \includegraphics[width=\textwidth]{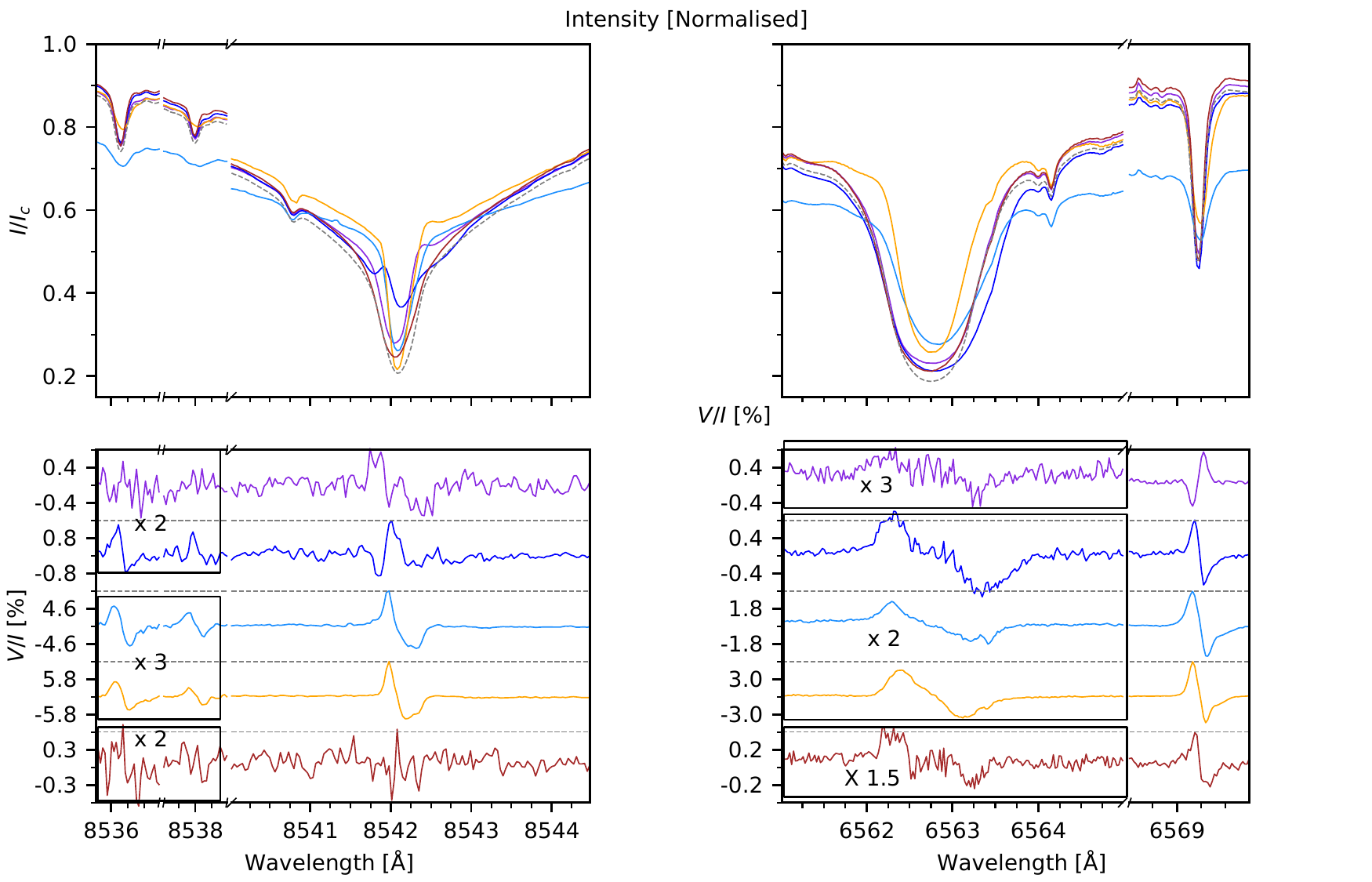}
    \caption{Sample \CaIR{} and \Halpha{} line profiles over one slit position located over the pore and surrounding region (marked by the brown dashed line in Fig.~\ref{fig:FOV}). The top two panels show the spectral images of Stokes~$I$ and $V$, respectively. A few selected profiles marked by colored dashed lines are plotted in the bottom two panels. The quiet-Sun profile (gray dashed) is also shown in the intensity plots for comparison. For better visibility, the amplitudes of a few Stokes~$V$ profiles are artificially multiplied by the factors as indicated in the respective panels.}
    \label{fig:castokes12}
\end{figure*}

\begin{figure*}[htbp]
    \centering
    \includegraphics[width=\textwidth,]{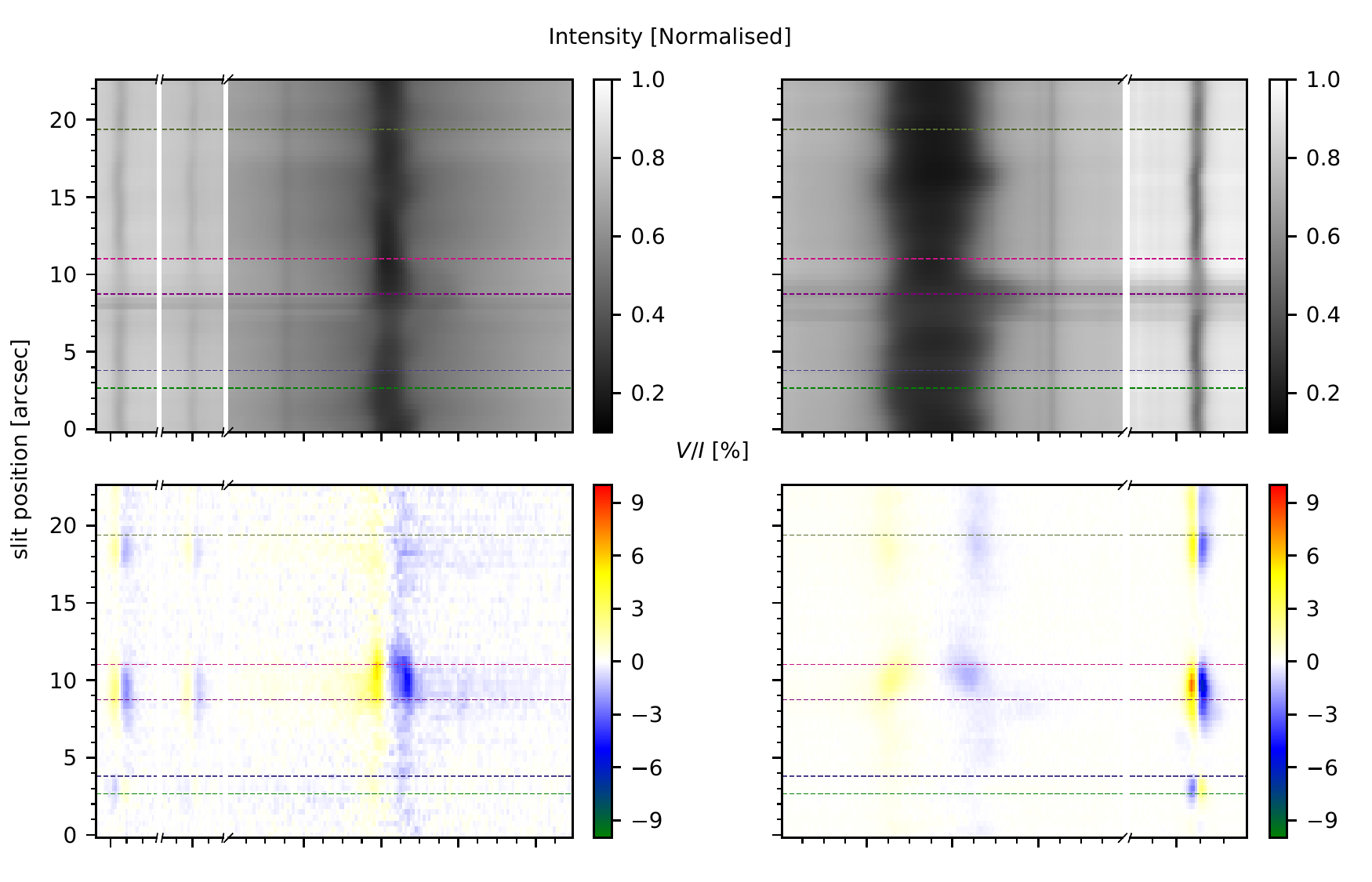}
    \includegraphics[width=\textwidth]{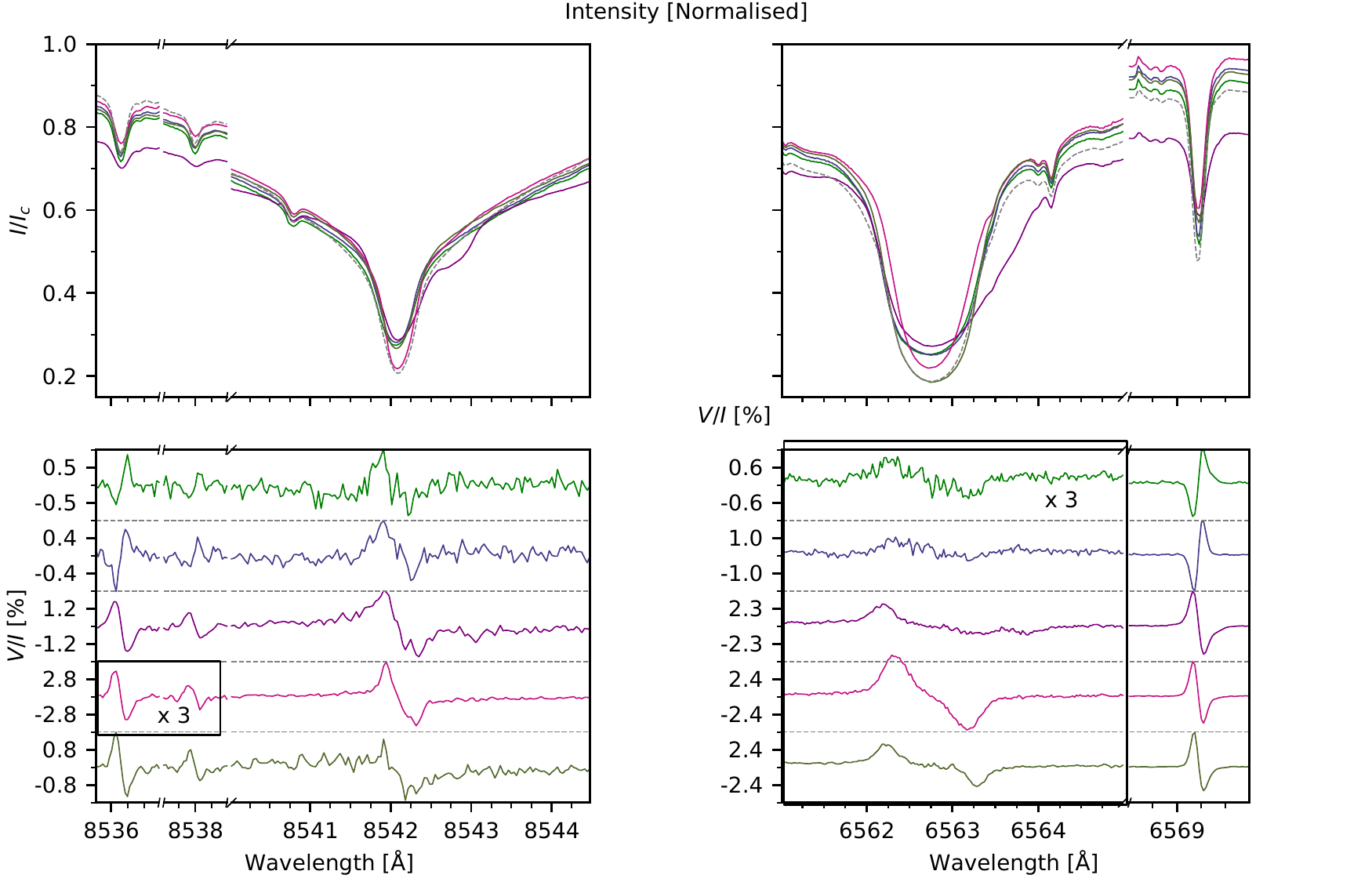}
    \caption{Same as Fig.~\ref{fig:castokes12} but for the slit passing through the region over negative polarity seen in \blos{} map inferred from ME inversions (see green line in Fig.~\ref{fig:FOV}). }
    \label{fig:castokes8}
\end{figure*}

\begin{figure*}[htbp]
    \centering
    \includegraphics[width=\textwidth]{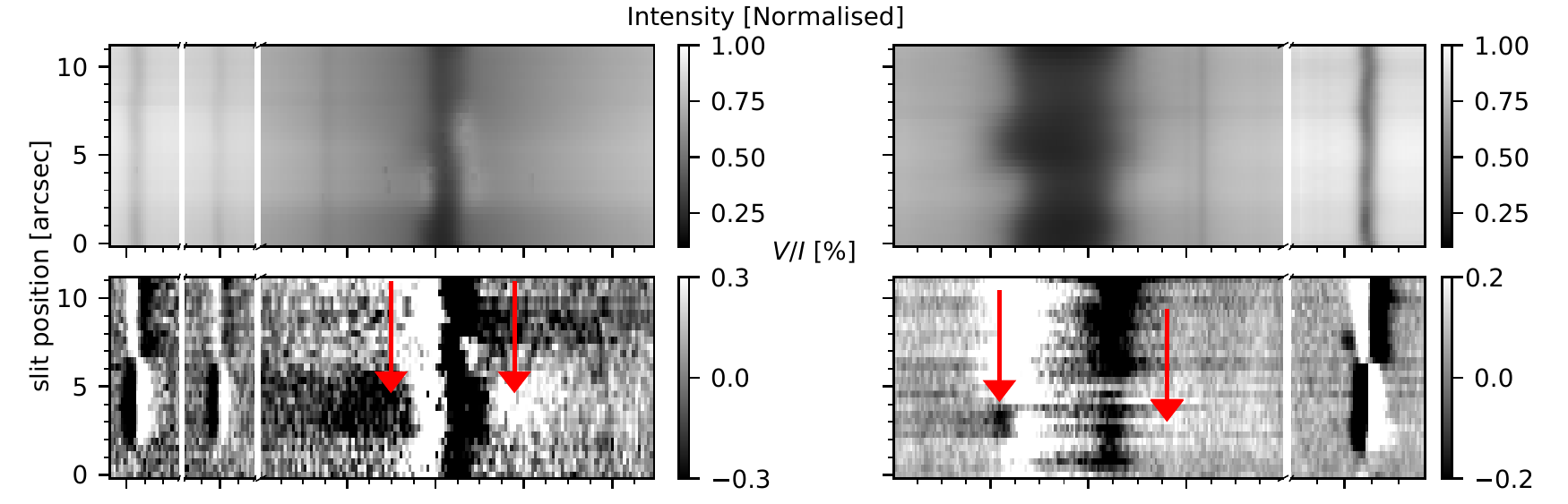}
    \includegraphics[width=\textwidth,keepaspectratio]{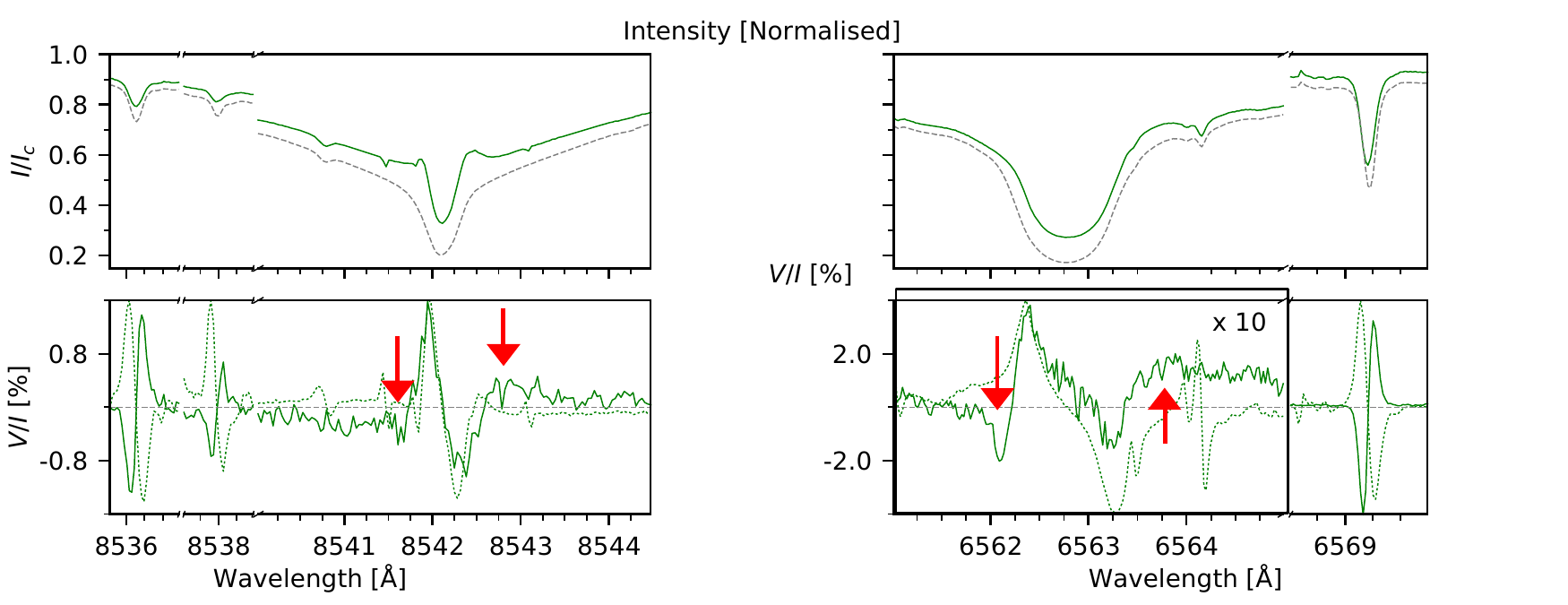}
    \caption{Sample \CaIR{} and \Halpha{} line profiles over a region (in another raster scan map) with opposite polarity in the photospheric and chromospheric lines. The top two panels show the spectral images of Stokes~$I$ and $V$, respectively. An average Stokes~$I$ and $V$ profile over the opposite polarity region is shown in the bottom two panels. The arrow indicates the wavelength position of the peak amplitude of the polarity reversal. The dotted green profile shows the derivative of Stokes~$I$ multiplied by $-$1. The quiet-Sun profile (gray dashed) is also shown in the intensity plots for comparison.  For better visibility, the amplitudes of the Stokes~$V$ profile of the \Halpha{} line are artificially multiplied by the factors as indicated in the respective panel.}
    \label{fig:castokes13}
\end{figure*}

The quiet-Sun profiles for the \Halpha{} and the \CaIR{} lines were calculated by averaging profiles of a few pixels in a region away from the pore with negligible signal in the Stokes~$V$ profiles.
As explained in appendix~\ref{sect:datareduc}, the wavelength calibration was done by comparing the quiet-Sun profile with the BASS 2000 atlas profile.
The \CaIR{} and \Halpha{} data were also corrected for spectral veil.

In the following paragraphs, we discuss a few selected profiles which are chosen such that they show distinct features from one another corresponding to various dynamics observed in the FOV.
In the top two panels of Fig.~\ref{fig:castokes12} we present sample spectral images of Stokes~$I$ and $V$ corresponding to a slit position marked by a brown line in Fig.~\ref{fig:FOV}.
In the bottom two panels of Fig.~\ref{fig:castokes12}, we show a few profiles from selected spatial locations marked using colored horizontal lines in the top two panels.
The left and right columns of panels correspond to the \CaIR{} and \Halpha{} spectrum, respectively.
%

There is a hint of enhancement in the intensity in the red wing of the \CaIR{} line profile at 8542.38~\AA\, ($\Delta v = +10.31$~\kms{}), at slit position near $\sim4\farcs2$ with the line core position (8542.04~\AA) slight blue-shifted ($\Delta v = -$1.5~\kms{}) compared to the quiet-Sun profile (see violet-colored profile).
The corresponding \Halpha{} profile shows nominal absorption.
%
The violet-colored profile lies in the opposite polarity region.
The Stokes~$V$ profiles of the \CaIR{} and \Halpha{} lines have weak signal with sign opposite to that of the Stokes~$V$ profile of the \FeIVISline{}.
There is hardly any signal in the Stokes~$V$ profile of the \SiIIRline{} and \FeIIRline{} lines that is above the noise level. This is because, as noted above, the SNR in \CaIR{} spectrum is about 5 times lower compared to that of \Halpha{} spectrum.

The intensity profile of the \CaIR{} line at slit position near $7\farcs8$ (blue-colored profile) shows emission in the blue wing  at 8541.9~\AA\, ($\Delta v = -$6.56~\kms{}) and line core (8542.12~\AA) redshifted ($\Delta v = +$1.4~\kms{}), signature of surge flow (more details are discussed in Nagaraju et al. under prep.).
The corresponding \Halpha{} and the wings of \CaIR{} intensity profiles also show red excursion compared to the quiet-Sun profile.
The Stokes~$V$ profile of the \CaIR{} line shows a sign reversal compared to the Stokes~$V$ profile of the \FeIVISline{} which is not due to change in the polarity of the magnetic field but due to emission in the blue wing \citep{1997A&A...324..763S}.
Such emission feature can only be caused by a change in the gradient of the source function, which also affects the sign of the Stokes~$V$ signal.

A sample profile over the pore is shown in cyan color.
%
%
%
%
The \CaIR{} and \Halpha{} lines show asymmetric Stokes~$V$ profiles with blue and red lobe amplitudes of ($\sim$9\%, $\sim$6\%) and ($\sim$1.3\%, $\sim$0.9\%), respectively.
In contrast, the \SiIIRline{}, \FeIIRline{} and \FeIVISline{} lines show relatively symmetric Stokes~$V$ profiles of amplitudes $\sim$1.6\%, $\sim$1\% and $\sim$3.5\% respectively.

%
There is a hint of enhancement in the red wing of the \CaIR{} line profile at the edge of the pore (yellow colored profile).
The Stokes~$V$ profiles of the \CaIR{} and \Halpha{} lines (similar to cyan-colored profile) show amplitude asymmetry.
This could be because of the presence of multiple Stokes components within one resolution element \citep{2000ApJ...544.1141S} or gradients in LOS velocity and magnetic field \citep[][]{2002ApJ...576.1048S}.

The brown-colored profile is an example of a quiescent profile away from the pore region.
%
%
The Stokes~$V$ amplitudes of the \CaIR{} and \Halpha{} line profiles are similar (0.2\%).

Fig.~\ref{fig:castokes8} shows spectropolarimetric images and spectral profiles for the slit position shown in green color in Fig.~\ref{fig:FOV} which passes over a location with stronger field within the negative polarity region and the pore edge.
The profiles over the negative polarity region are shown in green and purple colors.
The Stokes~$V$ amplitudes of the \SiIIRline{} and \CaIR{} lines are $\sim$0.7\%.
The green and purple colored intensity profiles of the \CaIR{} line are blueshifted by about $\Delta \lambda = -0.04$ and $-$0.02~\AA\, ($\Delta v = -1.4$ and 0.7~\kms{}), respectively, whereas there is no Doppler shift seen in the \SiIIRline{} and \FeIIRline{} lines.
%
%
The sign of the Stokes~$V$ profiles of the \SiIIRline{}, \FeIIRline{} and \FeIVISline{} lines is opposite to that of the sign of the Stokes~$V$ profiles of the \CaIR{} and \Halpha{} lines  which suggest that the Stokes~$V$ profile of the \Halpha{} line probes the chromospheric magnetic field.
This most probably is because the \Halpha{} and \CaIR{} line cores are sampling the canopy fields extending from a nearby region and overlying the opposite polarity region. 
To further demonstrate the existence of such field configuration, we show in Fig.~\ref{fig:castokes13} sample spectral images and profiles of the same region but from a raster scan recorded at a later time (16:08 UT). 
As clearly seen in this figure that the sign of the Stokes~$V$ profiles corresponding to the core of \CaIR{} and \Halpha{} lines are opposite to that of Stokes~$V$ profiles of the photospheric lines viz., \SiIIRline{}, \FeIIRline{}, and \FeIVISline{}(see spectral images from 2$\arcsec$ to 6$\arcsec$).
On the other hand, the Stokes profiles in the wings of the \CaIR{} and \Halpha{} lines show the polarity the same as that of the photospheric lines. 
This is consistent with the canopy field scenario as explained above since the line wings of \Halpha{} and \CaIR{} lines form deeper in the solar atmosphere, and hence the Stokes~$V$ profiles have the same sign as that of the photospheric lines.
However, as noted before, under weak field conditions, the Stokes~$V$ profiles change sign when the spectral features change from absorption to emission or vice-versa.
In order to make sure that the change in sign of Stokes~$V$ profiles from core to wing of the chromospheric lines is actually because of the change in polarity of the magnetic field but not due to change in spectral features, we have over-plotted the first derivative of Stokes~$I$ ($\frac{dI}{d\lambda}$: the dotted green curves in the bottom panel of Fig.~\ref{fig:castokes13} with the sign changed) over the Stokes~$V$ profiles. 
The reason for doing this is that under weak field conditions, Stokes~$V$ profile resembles $\frac{dI}{d\lambda}$ profile (see section~\ref{sect:wfa}).
The comparison between the Stokes~$V$ and $\frac{dI}{d\lambda}$ profiles clearly demonstrate that the Stokes~$V$ sign change from core to wings of the chromospheric lines is due to the change in the polarity of the magnetic field but not due to change in the emission or absorption features of Stokes~$I$.

The profile shown in maroon color (Fig.~\ref{fig:castokes8}) is another example of a profile in pore region.
There is a red excursion in the \CaIR{} line wings and \Halpha{} line core profile corresponding to a surge flow.
The Stokes~$V$ profiles of the \CaIR{}, \Halpha{}, and \FeIVISline{} lines show amplitude asymmetry, and the blue lobe show a positive sign suggesting positive polarity of the \blos{}.

The pink and khaki-colored profiles show profiles away from the pore region.
The intensity profiles of the \CaIR{} and \Halpha{} lines are similar to that of the quiet-Sun profile.
Amplitude asymmetry is present in the Stokes~$V$ profiles of the \CaIR{} and \Halpha{} lines.

\label{sect:methods}
\section{Methods}

\subsection{Weak field approximation} \label{sect:wfa}

%

The magnetic field from the \Halpha{} spectral line was inferred using the WFA. Under WFA the Stokes~$V$ is linearly related to \blos{} and ($\partial I / \partial \lambda$) through \citep{2004ASSL..307.....L}
\begin{equation}
\label{eq:wfa}
    V(\lambda) = - \Delta \lambda_{\mathrm{B}}\;\bar{g}\cos \theta\;\frac{\partial I}{\partial \lambda},
\end{equation}

and 

\begin{equation}
    \Delta \lambda_{\mathrm{B}} = 4.67 \times 10^{-13} \lambda_{0}^{2} B,
\end{equation}

where \deltab{} is expressed in \AA\,, $B$ in Gauss, $\bar{g}$ is effective Landé factor, $\theta$ is the inclination of $B$ with respect to the LOS, and $\lambda_{0}$ is the central wavelength of the spectral line (expressed in \AA).

The \blos{} can be derived from Eq.\ref{eq:wfa} using the linear regression formula \citep[e.g.,][]{2009ApJ...700.1391M},

\begin{equation}
    B_{\mathrm{LOS}} = - \frac{\Sigma_{\lambda} \frac{\partial I_{\mathrm{\lambda}}}{\partial \lambda} V(\lambda)}{C \Sigma_{\mathrm{\lambda}} (\frac{\partial I_{\mathrm{\lambda}}}{\partial \lambda})^{2}} \textcolor{red}{,}
\end{equation}
where $C=4.66 \times 10^{-13}\;\bar{g}\;\lambda_{\mathrm{0}}^{2}$.

We have used $\bar{g}=1.048$ following the investigation done by \citet{1994A&A...291..668C}.
We derived three values of \blos{}, one from the line core (\Halpha{}$\pm$0.35~\AA), the line wings ($[-1.5, -0.6]$ and $[+0.6, +1.5]$~\AA\,), and over the full \Halpha{} spectral line (\Halpha{}$\pm$1.5~\AA).
The spectral blends listed in Table~\ref{tab:wfaexclude} were excluded while calculating \blos{} using the WFA, as have done by \citet{2022A&A...659A.179J} and \citet[][]{2020JApA...41...10N}.
The WFA is applied on the \CaIR{} line within the wavelength range $\lambda_{0}\pm0.25$~\AA\, and the inferred \blos{} map is shown in panel (e) of Fig.~\ref{fig:FOV}.
The uncertainties in the values of \blos{} inferred from applying the WFA to the \CaIR{} and \Halpha{} data were estimated to be 23 and 18~G, respectively.

\begin{table}[htbp]
    \centering
    \begin{tabular}{ccc}
         \hline\hline
         $\lambda_{\mathrm{0}}$ [\AA] & $\Delta\lambda_{\mathrm{0}}$ [\AA] & Line\\
         \hline
         6562.44 & 0.05 & \ion{V}{2}\\
         6563.51 & 0.15 & \ion{Co}{1}\\
         6564.15 & 0.35 & Unknown\\
         \hline
    \end{tabular}
    \caption{The first and second column define the spectral blends removed before applying the WFA: $\lambda_{\mathrm{0}}\pm\Delta\lambda$. The third column indicates the element of the transition in case it is known.}
    \label{tab:wfaexclude}
\end{table}
\subsection{Milne-Eddington inversion}
We performed Milne-Eddington (ME) inversions \citep[see chapter 11 of][]{2007insp.book.....D} of the \FeIVISline{} data to infer the \blos{} (panel (f) of Fig.~\ref{fig:FOV}) utilizing pyMilne code, a parallel C++/Python implementation\footnote{\url{https://github.com/jaimedelacruz/pyMilne}} \citep{2019A&A...631A.153D}.
The \blos{} from the \Halpha{} line is inferred from the WFA as explained in section~\ref{sect:wfa}.
The stratification of the \blos{} from the photosphere to the chromosphere is inferred using simultaneous multi-line non-LTE inversions explained in section~\ref{sect:methodinv}.

\subsection{Non-LTE inversion}\label{sect:methodinv}

\begin{table*}[htbp]
\caption{Node positions (\logtau{} scale) used for inversions of different categories of profiles in \temp{}, \vlos{}, \vturb{}, and \blos{}} 
\label{table:inversionstrategy}
\begin{center}
\begin{tabular}{ c c c c  }
\hline\hline
Parameters & Category & Cycle 1 & Cycle 2\\
\hline
\temp{} & Quiescent & $-$5.5, $-$4,5, $-$3.5, $-$2.5, $-$1.5, 0 & $-$5.5, $-$4.5, $-$3.5, $-$2.5, $-$1.5, 0\\
& Emission & $-$4.8, $-$3.8, $-$2.9, $-$1.8, $-$0.9, 0 & $-$4.8, $-$3.8, $-$2.9, $-$1.8, $-$0.9, 0\\
\vlos{} & Quiescent & $-$4.5, $-$1 & $-$4.5, $-$1\\
& Emission & $-$6, $-$4.5, $-$3, $-$1 & $-$6, $-$4.5, $-$3, $-$1\\
\vturb{} & All & $-$5,$-$4, $-$3, $-$1 & $-$5,$-$4, $-$3, $-$1\\
\blos{} & All & -- & $-$4.5, $-$1\\
\hline
\end{tabular}
\end{center}
\end{table*}

The MPI-parallel STockholm inversion Code \citep{2019A&A...623A..74D, 2016ApJ...830L..30D} is used to retrieve the stratification of atmospheric parameters.
STiC is based on a modified version of the RH radiative transfer code \citep{2001ApJ...557..389U} and solves the polarised radiative transfer equation using cubic Bezier solvers \citep{2013ApJ...764...33D}.
In non-LTE, assuming statistical equilibrium, it can fit multiple spectral lines simultaneously.
It employs the fast approximation to account for partial re-distribution effects (PRD) \citep[for more details][]{2012A&A...543A.109L}.
STiC assumes plane-parallel geometry to fit the intensity in each pixel (also called the 1.5D approximation).
STiC uses an LTE equation-of-state obtained from the library functions in the Spectroscopy Made Easy (SME) package code \citep{2017A&A...597A..16P}.
The optical depth scale at 5000~\AA\, (500~nm), abbreviated \logtau{}, is used to stratify atmospheric parameters.

We have inverted the Stokes~$I$ and $V$ profiles of the \CaIR{}, \SiIIRline{} and \FeIIRline{} lines simultaneously to infer the stratification of temperature (\temp{}), LOS velocity (\vlos{}), microturbulence (\vturb{}) and LOS magnetic field (\blos{}).
We used a 6-level \ion{Ca}{2} atom.
The \CaHK profiles were modeled in PRD \citep{1974ApJ...192..769M, 1989A&A...213..360U}, and \ion{Ca}{2}~IR lines were synthesized in complete re-distribution (CRD) approximation.
The atomic parameters of the \SiIIRline{} and \FeIIRline{} lines were obtained from the Vienna Atomic Line Database (VALD3) \citep{2015PhyS...90e4005R}, and Kurucz's line lists \citep{2011CaJPh..89..417K}, respectively and synthesized under LTE approximation.
The upper level of the \SiIIRline{} transition was treated with $J_{1}$-$l$ ($JK$) coupling scheme as described in appendix~\ref{sect:jk}.
The latest version of the STiC inversion code has been upgraded to allow for the treatment of atomic levels of Kurucz's lines in $JK$ coupling.
An empirical $\log \mathrm{gf}$ value of $-1.4$ was used for the \FeIIRline{} transition \citep{2007ApJS..169..439S}.

We used the $k$-means clustering to group the Stokes~$I$ profiles in different clusters such that similar-shaped profiles were grouped in one cluster.
We then inverted the mean profile of each of those clusters to derive the stratification of \temp{}, \vlos{}, and \vturb{}.
Finally, the inferred stratification was used as the initial guess atmosphere to infer the stratification of the atmospheric parameters of the FOV, similar to the approach used by \citet{2021A&A...655A..28N} and \citet{2022A&A...668A.153M}.
In the second cycle, we used the values of \blos{} derived from the ME inversions from the \FeIVISline{} line and WFA of the \CaIR{} line as the guess values of the \blos{} at \logtau{} = $-$1 and $-$4.5, respectively.
Earlier studies have found that the Stokes~$V$ profiles of the \CaIR{} line have maximum response to the perturbations in the \blos{} between \logtau{} = $-$4 and $-$5 \citep[][]{2016MNRAS.459.3363Q, 2018A&A...619A..63J, 2019ApJ...873..126M}. 
Table~\ref{table:inversionstrategy} describes the node positions used for different categories of profiles.
Quiescent profiles are nominal absorption profiles (910 profiles), and emission profiles are profiles that have an emission (or a hint of emission) in either blue or red or both the wings of the \CaIR{} line (110 profiles).
The quality of inversion fits are discussed in appendix~\ref{sect:quality}. 

We set the average velocity in the pore region in the photosphere (\logtau range of [$-1$, $0$]) to rest for the absolute velocity calibration.
%
%
%
With respect to the average profile in the pore, the quiet-Sun profile is blue-shifted, that is, after velocity calibration, the \vlos{} stratification inferred from the quiet-Sun profile shows an upflow of about $-$3~\kms{}.

\section{Results and discussion} \label{sec:results}

\subsection{Results from the WFA and ME inversions}

The \blos{} map inferred from the ME inversion of Stokes~$I$ and ~$V$ profiles of \FeIVISline{} line is shown in panel~(f) of Fig.~\ref{fig:FOV}. The maximum \blos{} strength found is +800~G. The panel~(e) of Fig.~\ref{fig:FOV} shows the \blos{} map inferred from WFA applied on the \CaIR{} within the wavelength range of $\lambda_0\pm 0.25$~\AA\, and the maximum \blos{} field strength found is +600~G. The region in the photosphere with opposite polarity of the \blos{} with respect to the pore has a field strength of about $-$200~G which is absent in the chromospheric \blos{} map.
%

\subsection{Non-LTE inversion results}

\begin{figure*}
    \centering
    \includegraphics[width=\textwidth]{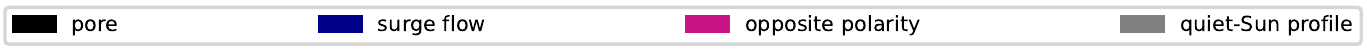}
    \includegraphics[width=\textwidth]{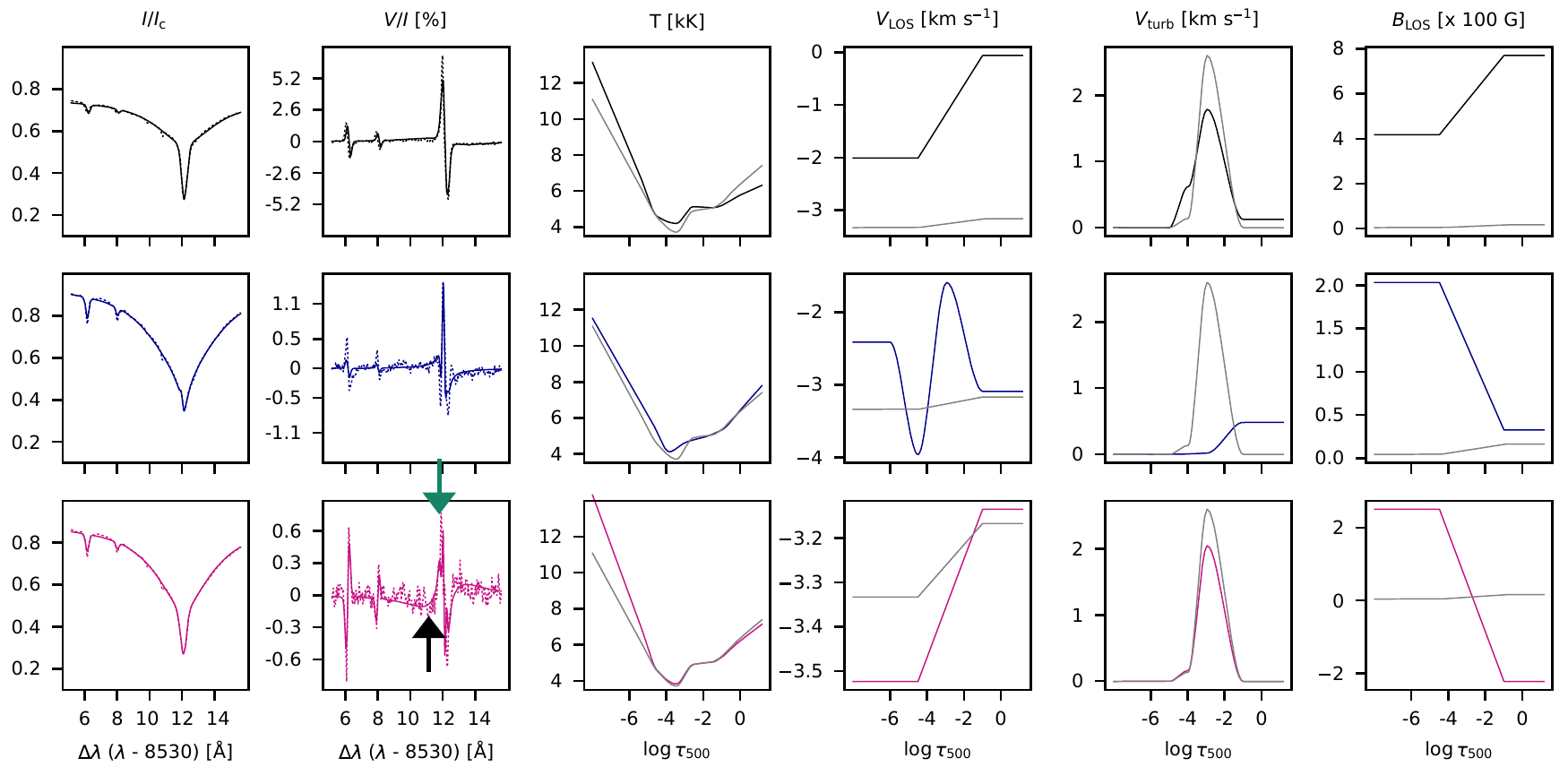}
    \caption{Inversion results of a few average profiles from different regions of the FOV. The dotted and dashed curves in the first two columns show the observed and fitted Stokes~$I$ and $V$ profiles, respectively. The next three columns show the stratification of the \temp{}, \vlos{}, \vturb{} and \blos{} inferred from the inversions. The gray-colored curve shows the stratification of atmospheric parameters inferred from the inversions of an average quiet-Sun profile. The black and green arrows indicate the wavelength position in the wing and near the core of the \CaIR{} line, respectively.}
    \label{fig:individualinversion}
\end{figure*}

\begin{figure}
    \centering
    \includegraphics[width=0.4\textwidth]{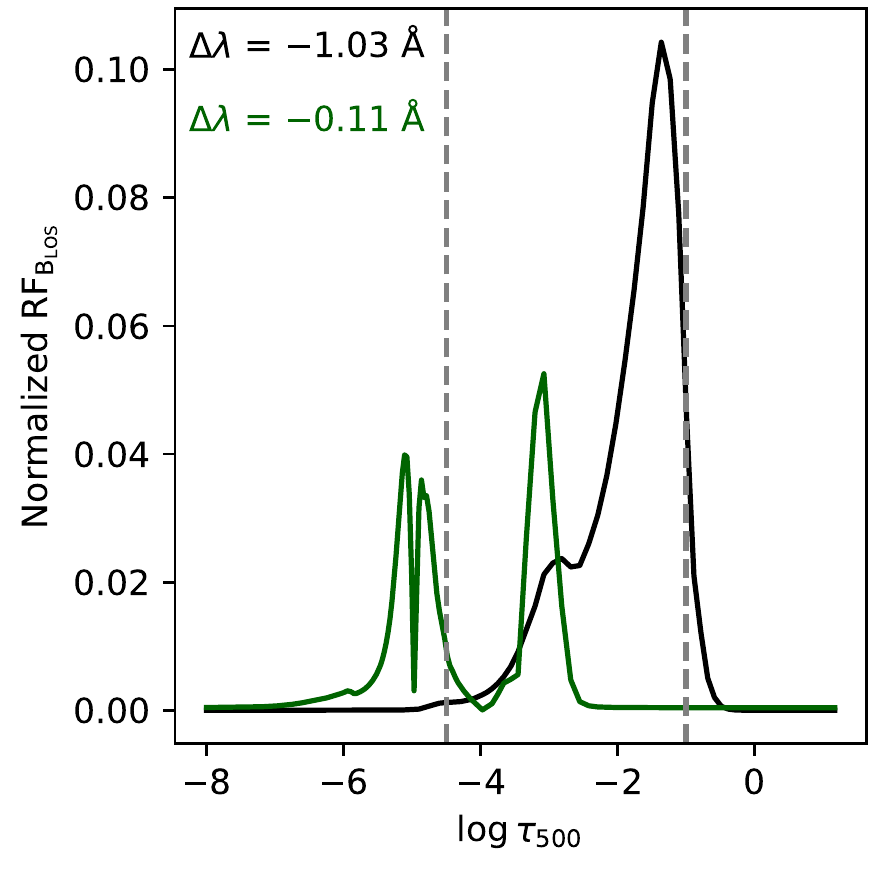}
    \caption{Normalized response function to the perturbations in the \blos{} at the wavelength positions $\Delta \lambda$ = $-$1.03 and $-$0.11~\AA\, (marked by black and green arrows in Fig.~\ref{fig:individualinversion}) from the \CaIR{} line core. The node positions for the \blos{} used in the inversions are represented by the vertical dashed gray lines. }
    \label{fig:opppolresp}
\end{figure}

\begin{figure}[htbp]
    \centering
    \includegraphics[height=8.26in]{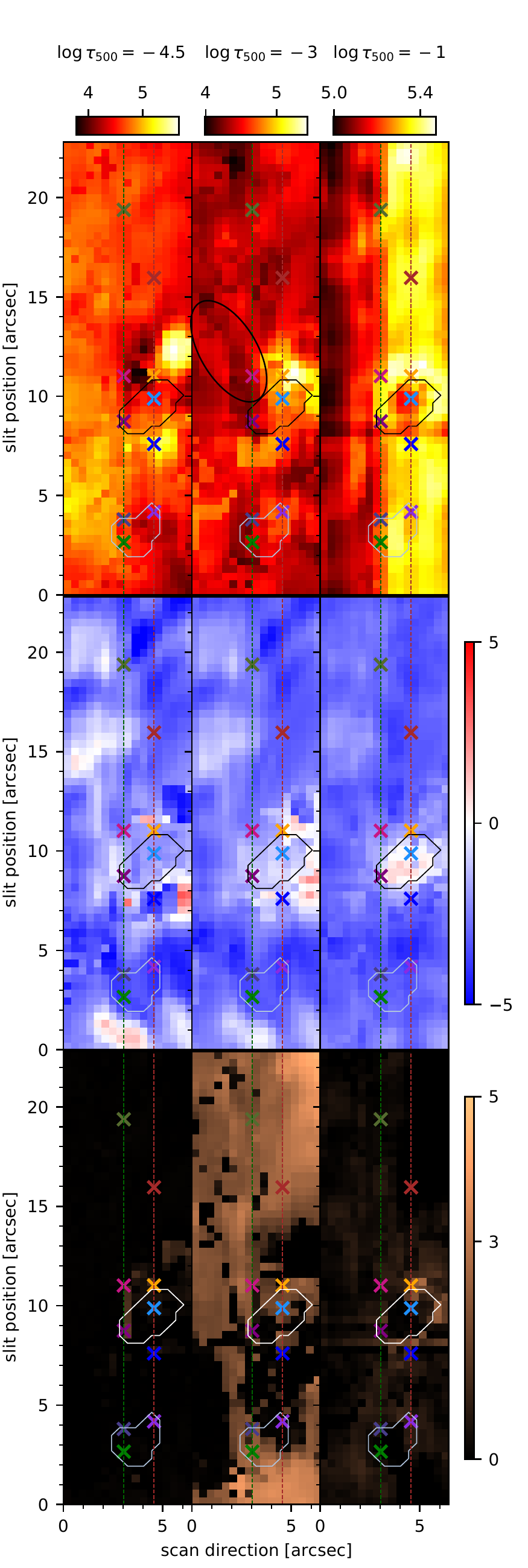}
    \caption{Maps of \temp{}, \vlos{} and \vturb{} (row-wise) at \logtau{} = $-$4.5, $-$3, $-$1 (column-wise) inferred from the inversions of the FOV. The contours, similar to Fig.~\ref{fig:FOV}, show the location of the pore and the opposite polarity region. The slanted elliptical contour in the middle panel of the \temp{} maps show the location of the dark fibrillar region seen in the panel (d) of Fig.~\ref{fig:FOV}.}
    \label{fig:inversion}
\end{figure}

\begin{figure*}[htbp]
    \centering
    \includegraphics[height=8.26in]{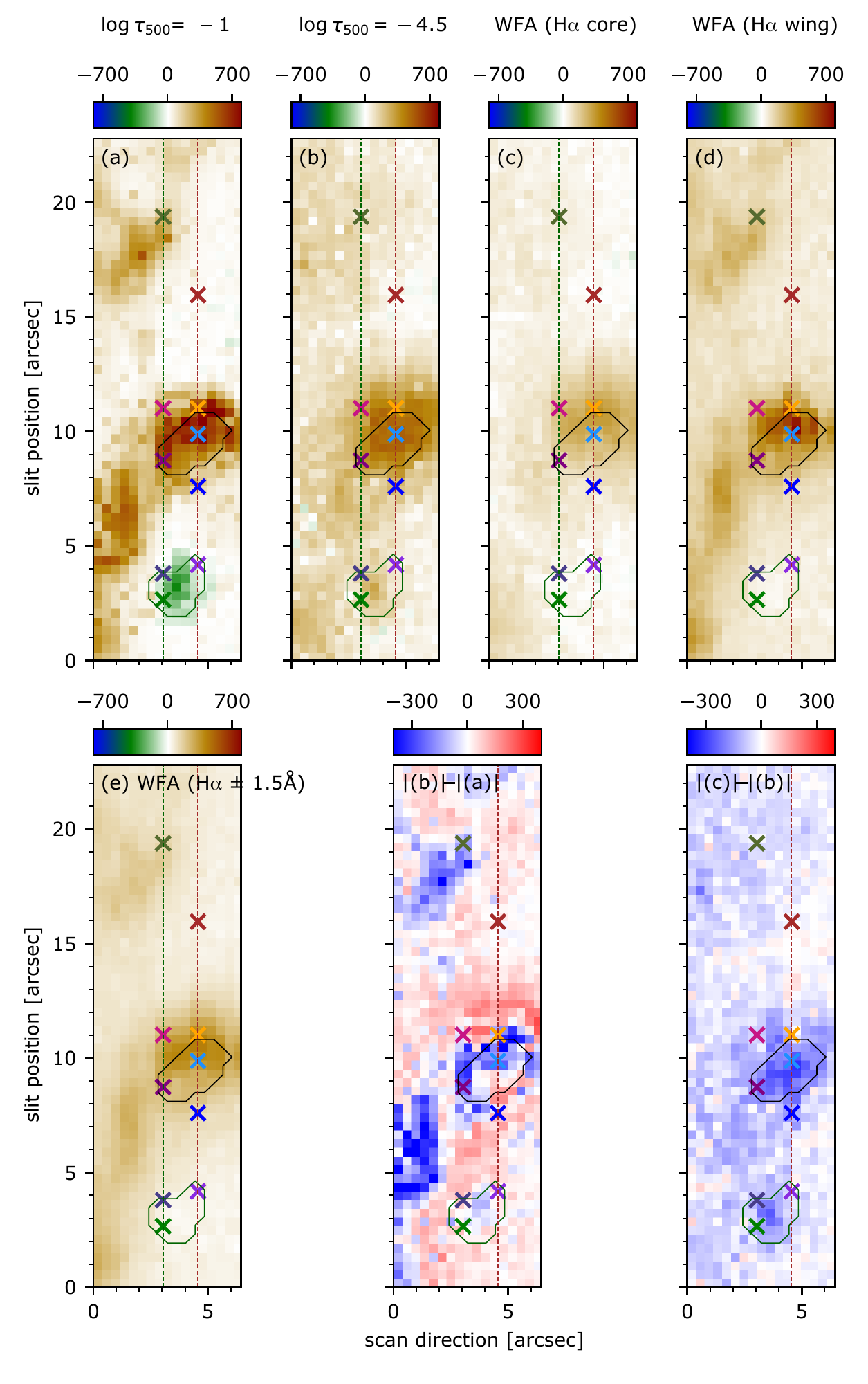}
    \caption{Maps of the \blos{} inferred from the inversions and the WFA. Panels (a) and (b) show the maps of \blos{} at \logtau{} = $-$1 and $-$4.5, respectively. Panel (c) show the \blos{} inferred from the WFA of \Halpha{} spectral line in the spectral range $\pm0.35$~\AA\,, panel (d) show the WFA inferred from the wings of \Halpha{} spectral line in the range [$-$1.5, $-$0.6] and [$+$0.6, $+$1.5]~\AA\,, and panel (e) show the map of \blos{} from WFA inferred from \Halpha{}$\pm$1.5~\AA\,. Change in the stratification of $|$\blos{}$|$ is shown in the rightmost two panels of the bottom row as indicated on each panel. The black and green contours show the location of the pore and opposite polarity region, respectively.}
    \label{fig:magfield}
\end{figure*}

In this section we discuss about the results from the inversion of Stokes~$I$ and ~$V$ profiles of the \CaIR{}, \SiIIRline{} and \FeIIRline{} lines using the STiC inversion code.

In Fig.~\ref{fig:individualinversion} we show inversion results of profiles averaged (to increase SNR) from a 3$\times$3 box about the selected regions of the FOV, viz., pore, surge flow, and region with opposite polarity of the magnetic field.
The average profile for the pore is calculated by averaging the pixels about the darkest pixel in the pore, for the surge flow profile by averaging about the blue-colored profile in Fig.~\ref{fig:castokes12} and for the opposite polarity profile by averaging about the green colored profile in Fig.~\ref{fig:castokes8}.

The \temp{} stratification inferred from the pore profile has lower value of \temp{} at the photospheric layers (\logtau{}$\sim-$1) and higher values at the chromospheric layers (\logtau{}$\leq-$4) compared to that inferred from the inversion of the quiet-Sun profile.
The \vlos{} at photospheric layers (\logtau{}$\sim-$1) is zero (calibrated) and at the chromospheric layers (\logtau{} = $-$4.5) show upflow of about $-$2~\kms{}.
The \vturb{} is non-zero between \logtau{} = $-$1 and $-$4.
The value of \blos{} at the photospheric and chromospheric layers is about $+$800~G and $+$400~G, respectively.

The reversal of the sign of the Stokes~$V$ profile of the \CaIR{} line shown in Fig.~\ref{fig:individualinversion} corresponding to surge flow region is a result of emission feature in the blue wing of the \CaIR{} line and is not an indication of any change in polarity of the \blos{}.
Accordingly the \temp{} show enhancement of about 600~K compared to \temp{} inferred from the quiet-Sun profile at \logtau{} = $-$4.5 simultaneous with an upflow of about $-$4~\kms{}.
For the pixels having spectral profiles similar to the surge flow, i.e., an emission feature in the blue wing and a red excursion in the Stokes~$I$, the \vturb{} is small at all depth positions of \logtau{}.
It may be possible that such low values of \vturb{} may have some contribution in the higher temperature values inferred for the pixels in the surge flow region.
\citet[][]{2022A&A...659A.165D} have shown that there is a degeneracy between the \temp{} and the \vturb{}.
When the spectral profiles are in emission, the \temp{} and \vturb{} are anti-correlated; that is, an increase in \temp{} broadens the spectral line, and thus to maintain the same width, the \vturb{} must be decreased.
However, the enhancement in \temp{} and minimal values of \vturb{} are necessary to achieve a satisfactory fit of the emission feature \citep[for more discussion see][]{2022A&A...668A.153M}.
%
%
%
%
The value of \blos{} at the photospheric and chromospheric layers is about $+$50~G and $+$200~G, respectively.
The reason for the increased value of the magnetic field in the chromosphere compared to the photosphere is because the line core of the \CaIR{} line is sampling the canopy fields overlying the pore and the nearby opposite polarity region.
%

The \temp{}, \vlos{} and \vturb{} stratification inferred from the opposite polarity profile is similar to that inferred from the quiet-Sun profile.
The value of \blos{} at the photospheric and chromospheric layers is about $-$220~G and $+$250~G, respectively, again suggesting a canopy structure.

As described in section~\ref{sect:observations}, above the positive polarity region, the sign of the Stokes~$V$ profile in the wings of the \CaIR{} and \Halpha{} lines is opposite to that of in the core.
This is probably because the line cores of the \Halpha{} and \CaIR{} lines are sampling the canopy fields overlying the opposite polarity region.
To verify that the wings and the core of the \CaIR{} line are indeed sampling different layers of the atmosphere, in Fig.~\ref{fig:opppolresp} we show the response to perturbations of the \blos{} at one wavelength position in the wing ($\Delta \lambda$ = $-$1.03~\AA) and another one in the core ($\Delta \lambda$ = $-$0.11~\AA) of the line where the sign of the Stokes~$V$ is opposite.
%
%
\cite{1975SoPh...43..289B} define the response function (RF) for a physical parameter $X$ as $RF_{X}(\tau, \lambda)\;=\;\delta I(\lambda)/\delta X(\tau)$.
Response functions contain information on how the Stokes parameters at different wavelength positions are sensitive to perturbations of a physical parameter at different \logtau{}.
%
%
The response function in the line wing of the \CaIR{} line shows a dominant contribution of the photospheric fields (\logtau{}$\simeq-$1.5), in contrast, the response function near the line core shows a dominant contribution of the chromospheric fields (\logtau{} $\leq$ $-3$).


The maps of \temp{}, \vlos{} and \vturb{} inferred from the inversions of the FOV at \logtau{} = $-$4.5, $-$3 and $-$1 are presented in Fig.~\ref{fig:inversion}.
%
The morphological structure of the \temp{} map at \logtau{} = $-$1 is similar to the far wing image of the \CaIR{} line (see panel (a) of Fig.~\ref{fig:FOV}) with a decrease in \temp{} of about $300$~K in the pore compared to surrounding background.
There is a weak downflowing ($+$2~\kms{}) region near ($x$, $y$) = ($3\farcs5$, $8\farcs8$) while the rest of the pore does not show any signature of plasma flows.
The value of \vturb{} is about 0--1~\kms{} in the \vturb{} maps at \logtau{} = $-$1 except at a few regions where the \vturb{} is about 2--3~\kms{} such as the pore boundary (6$\arcsec$, $10\farcs8$).

The \temp{} at \logtau{} = $-$3 ranges from 4--5.5~kK with a region of higher temperature (\temp{}$\sim$5.4~kK) seen near the pore boundary (6$\arcsec$, $10\farcs8$).
The temperature near the negative polarity region (in ME \blos{} map) is about 4.2~kK.
The dark fibril-like lanes starting at (2$\arcsec$, 8$\farcs$8), indicated by the ellipse-shaped contour in Fig.~\ref{fig:inversion}, can also be seen in the \Halpha{} line core image (see panel (d) of Fig.~\ref{fig:FOV}).
The majority of the pore region show upflows of up to $-3$~\kms{}.
The \vturb{} inside the pore is about 0--3~\kms{} and outside the pore is about 3--5~\kms{}.

The morphological structure of the \temp{} map at \logtau{} = $-$4.5 looks similar to the \CaIR{} line core image (see panel (c) of Fig.~\ref{fig:FOV}).
The brightest and darkest pixels in panel(c) of Fig.~\ref{fig:FOV} correspond to \temp{} of about $5.7$~kK and $3.8$~kK, respectively.
In general, the FOV show upflows of up to $-$3~\kms{} with two small regions showing downflows of about $+$2~\kms{}.
The strong upflows up to $-$5~\kms{} are located in regions of higher temperature.
There is almost zero \vturb{} in the \vturb{} map in all regions of the FOV.

The panels (a) and (b) of Fig.~\ref{fig:magfield} show maps of the magnetic field at \logtau{} = $-$1 and $-$4.5.
The change in the stratification of $|$\blos{}$|$ inferred from inversions is shown in the middle panel of the bottom row of Fig.~\ref{fig:magfield}.
The morphological structure of the \blos{} map at the photospheric layers (at \logtau{} = $-$1, panel (a) of Fig.~\ref{fig:magfield}) is similar to panel (e) of Fig.~\ref{fig:FOV} with strong positive polarity in the pore ($\sim+$800~G) and nearby regions ($\sim+$350~G).
An opposite polarity with a magnitude of $\sim-$200~G is seen in the \blos{} map.
The field at \logtau{} = $-$4.5 has positive polarity with a maximum \blos{} of $\sim+600$~G in the pore and about $\sim+$300~G in the opposite polarity region.
In general, the structure of the \blos{} at \logtau{} = $-$4.5 is more spread out compared to that of at \logtau{} = $-$1 (see panel (b) of Fig.~\ref{fig:magfield}).
In addition, in the regions outside of the pore the \blos{} strength at \logtau{} = $-$4.5 has increased compared to that at \logtau{} = $-$1, suggesting a magnetic canopy-like structure (see middle panel in the bottom row of Fig.~\ref{fig:magfield}).
Using multiple spectral lines, the magnetic canopies around pores have also been reported by many authors in the recent literature \citep[for eg.][]{2022ApJ...930...87S, 2022cosp...44.2512T, 2019A&A...630A..86B, 2016ApJ...825...75M, 2012ApJ...747L..18S, 1996A&A...316..229K}.
The magnetic field strengths inferred at \logtau{} = $-$1 (the photosphere) and \logtau{} = $-$4.5 (the chromosphere) are comparable with the values reported by many authors in recent literature, who studied the stratification of \blos{} in pore using spectral lines of the \FeI{} atom, and \CaIR{} and \HeIIRline{} lines \citep{2022arXiv220211679S,2020IAUS..354...46N, 2019A&A...632A.112Y,2016MNRAS.460.1476Q,2015A&A...580L...1J,2013A&A...560A..84S,2012A&A...546A..26C}.

\subsection{Comparison of \blos{} inferred from the \CaIR{} and \Halpha{} lines}\label{sect:rescompare}

\begin{figure}[htbp]
    \includegraphics[width=0.5\textwidth]{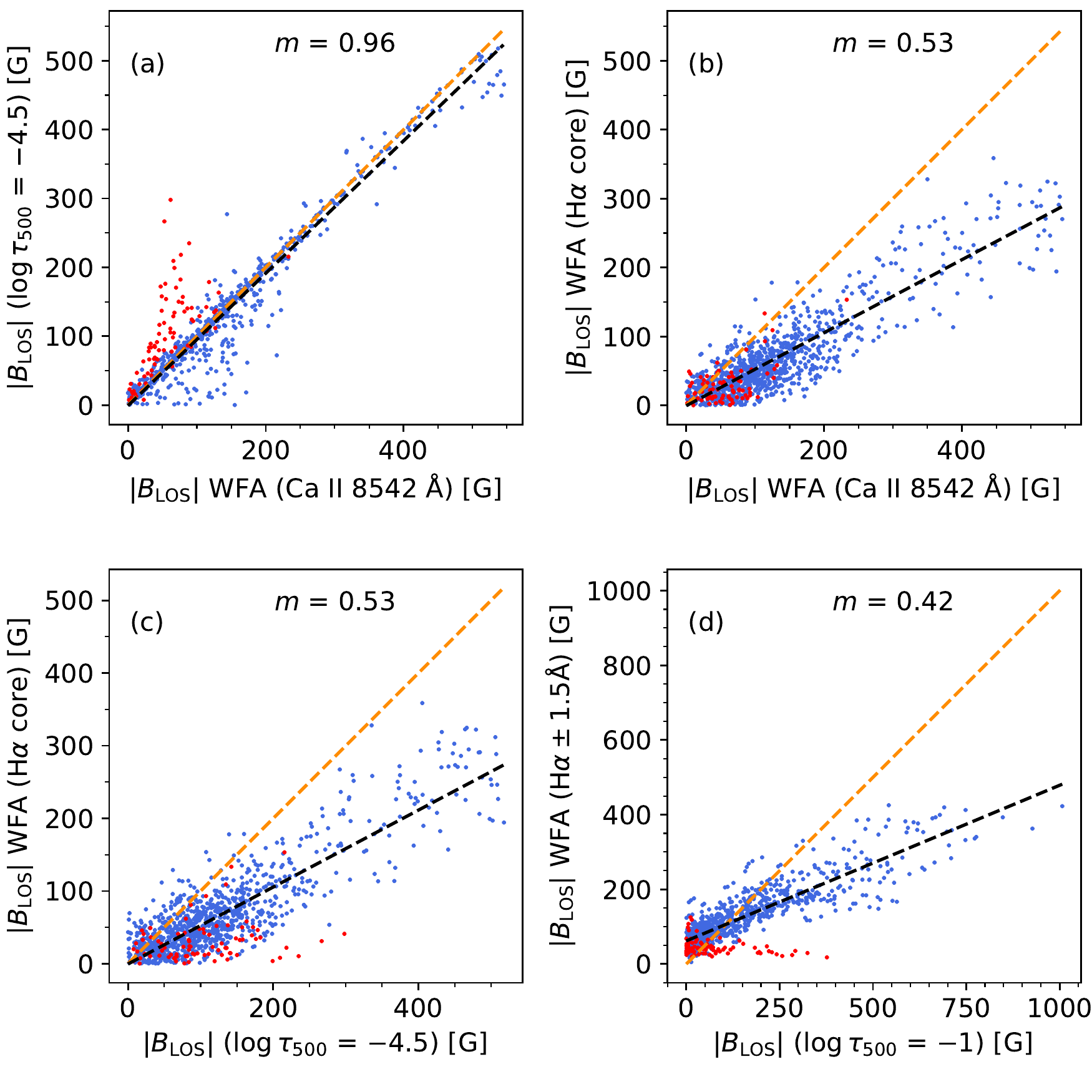}
    \caption{Comparison of the magnitude of \blos{} inferred from inversions and the WFA. The red and blue colors represent the pixels at photospheric layers with negative and positive polarity of the \blos{}, respectively. Panel (a) shows the comparison between the \blos{} inferred at \logtau{} = $-$4.5 with that inferred from WFA on the \CaIR{} line. Panels (b) and (c) show the comparison between the \blos{} inferred by applying the WFA on the \Halpha{} line core with the \blos{} inferred from the WFA on \CaIR{} line and at \logtau{} = $-$4.5, respectively. Panel (d) show comparison between the \blos{}  inferred by applying the WFA  on the \Halpha{}$\pm$1.5~\AA\, with \blos{} at \logtau{} = $-$1. The black colored line shows the linear fit whose slope is indicated by $m$. The fiducial line is shown in yellow color for comparison. }
    \label{fig:magscatter}
\end{figure}

The panels (c), (d) and (e) of Fig.~\ref{fig:magfield} show maps of the magnetic field inferred from the WFA method applied to the Stokes~$I$ and $V$ profiles of the \Halpha{} line.
The difference in amplitude of the $|$\blos{}$|$ inferred from the \Halpha{} line with that of inferred from the inversions is shown in the two rightmost panels of the bottom row of Fig.~\ref{fig:magfield}.
Figure~\ref{fig:magscatter} shows the scatter plots between the $|$\blos{}$|$ inferred from inversions and the WFA.

The field strengths inferred from inversions of the \CaIR{} line at \logtau{} = $-$4.5 are comparable to those of inferred from the WFA of the \CaIR{} line (see panel (a) of Fig.~\ref{fig:magscatter}), suggesting consistency between the two methods in inferring the \blos{}.
The magnetic field map inferred from the WFA of the \Halpha{} line core (\Halpha{}$\pm$0.35~\AA) has a similar morphological structure to the map inferred from inversions of \CaIR{} line at \logtau{} = $-$4.5 (see panels (b) and (c) of Fig.~\ref{fig:magfield}) suggesting the \Halpha{} line core probes the chromospheric magnetic field.
The \blos{} field strength inferred from WFA of \Halpha{} line core is almost half (0.53 times) of that of inferred from inversions at \logtau{} = $-$4.5 and the WFA of the \CaIR{} line (see the rightmost panel in the bottom row of Fig.~\ref{fig:magfield} and panel (b) and (c) of Fig.~\ref{fig:magscatter}).
This could be due to sensitivity of the core of the \Halpha{} line to the magnetic field in the higher atmospheric layers than that of \CaIR{} line. Or due to systematic underestimation of \blos{} using the WFA from the \Halpha{} line.
A detailed analysis of the \Halpha{} Stokes~$I$ and $V$ profiles synthesized using current state-of-the-art model atmospheres of active regions and quiet Sun, taking into account the fine-structure sub-levels ($nlj$) of the atomic levels ($n$),  3D radiative transfer and the Zeeman effect, is required.
The maps of \blos{} inferred from the WFA method on the wings and the full spectral range of the \Halpha{} line show a morphological structure similar to that of \logtau{} = $-$1 (see panel (d) and (e) of Fig.~\ref{fig:magfield}), suggesting a significant contribution from the photospheric fields.
However, the negative polarity region is not clearly seen.
This is because, as explained in section~\ref{sect:observations}, there are very few spectral pixels with a clear signal of opposite sign in Stokes~$V$ profile in the line core and wings of the \Halpha{} line.
%
%
The negative polarity region is very well reproduced through inversions because of good signal in Stokes~$V$ profiles of the \SiIIRline{} and \FeIIRline{} lines.
The $|$\blos{}$|$ inferred from the full spectral range of the \Halpha{} line is weaker by a factor of 0.42 than that of \blos{} at \logtau{} = $-$1, which is consistent with previous studies (see panel (d) of the Fig.~\ref{fig:magscatter}) \citep[][]{1971SoPh...16..384A, 2004ApJ...606.1233B, 2005PASJ...57..235H, 2008ApJ...678..531N}.


The above comparison of \blos{} inferred from the \Halpha{} and the \CaIR{} lines suggest that the line core of the \Halpha{} line is sensitive to the chromospheric magnetic fields while the wings and the full \Halpha{} line exhibit significant sensitivity to the photospheric magnetic fields.

\section{Conclusions}\label{sec:conclusions}

In this paper, we presented an analysis of spectropolarimetric observations recorded simultaneously in the \Halpha{} and \CaIR{} lines of a pore with positive magnetic polarity, as well as the surrounding region that covers also a negative polarity magnetic field region.
This is towards the goal of exploring the diagnostic potential of the \Halpha{} line to probe the chromospheric magnetic field.
To estimate \blos{} field from the \Halpha{} line we have used the WFA, and from \CaIR{} line we have used both the WFA and inversion methods.
The similarity between the magnetic field morphology inferred by applying the WFA on the core of the \Halpha{} and \CaIR{} lines and inferred from inversions at \logtau{}=$-$4.5 is a clear evidence that the \Halpha{} line core probes the chromospheric magnetic field.
This evidence is even more striking in the region above the negative polarity region in which the \Halpha{} and \CaIR{} line cores exhibit positive polarity where as the line wings exhibit negative polarity. 
This is because of the canopy fields from the dominant positive polarity region extended at the chromospheric heights overlying above the negative polarity region, which is mostly confined to photospheric heights.
It is found from the quantitative comparison that \blos{} fields estimated from the \Halpha{} line core are about $\approx 0.53$ times that of estimated from the \CaIR{} line core.
This may suggest that the magnetic sensitivity of the \Halpha{} line core is located in the higher layers in the solar atmosphere than that of the \CaIR{} line core.
However, there is a possibility that \blos{} values are systematically underestimated from the \Halpha{} line under WFA.
Further investigation through multi-line spectropolarimetric observations and 3D radiative transfer calculations are required to fully understand the diagnostic potential of the \Halpha{} line to probe the chromospheric magnetic field.

\begin{acknowledgments}
We are grateful to the anonymous referee for their valuable feedback, which helped to enhance the quality of this manuscript.
This research has made use of the High-Performance Computing (HPC) resources (NOVA cluster) made available by the Computer Center of the Indian Institute of Astrophysics, Bangalore.
This project has received funding from the European Research Council (ERC) under the European Union's Horizon 2020 research and innovation program (SUNMAG, grant agreement 759548).
This work has made use of the VALD database, operated at Uppsala University, the Institute of Astronomy RAS in Moscow, and the University of Vienna.
This research has made use of NASA’s Astrophysics Data System Bibliographic Services.
This research used version 3.1.6 \citep{sunpy_community2020} of the SunPy open source software package \citep{Barnes_2020}.
We have also used the packages h5py \citep{collette_python_hdf5_2014}, matplotlib \citep{Hunter2007} and numpy \citep{harris2020array} to carry out our data analysis.
\end{acknowledgments}

\vspace{5mm}
\facilities{SPINOR(DST), STiC}

\software{SunPy,
          NumPy,
          matplotlib,
          RH,
          STiC,
          pyMilne
          }

\bibliography{sample631}{}
\bibliographystyle{aasjournal}

\appendix

\section{Data reduction}
\label{sect:datareduc}
The data are reduced with standard procedures of bias and flat fielding.
Calibration data with the procedures described in \cite{2006SoPh..235...55S} are used to correct for instrumental polarization.
No absolute wavelength calibration is done because of the absence of suitable telluric lines.
Instead, we average a few spatial pixels in the quiescent region outside the pore (quiet-Sun profile) and fit the quiet-Sun profile over the full \CaIR{} spectral range (\imeanobs{}) with the BASS 2000 atlas \citep[][]{2009ASPC..405..397P}.

The procedures of the spectral veil correction, SI intensity calibration, and estimation of spectral Point Spread Function (PSF) for the \CaIR{} and \Halpha{} data are described as follows.
We followed the spectral veil correction process described in \citet{2016A&A...596A...2B} with an additional step correcting the tilt in the spectrum continuum.
The continuum of the raw data, because of detector flat-field residuals and pre-filter shape, is tilted.
These tilts were corrected in the data reduction pipeline by subtracting a linear fit ($y=a+b\lambda$) with the average spectrum.
However, the observed spectral range is not symmetric with respect to the core of the \CaIR{} and \Halpha{} lines, and there is an inherent tilt present even in the BASS 2000 atlas.
Hence we again corrected this over-correction in the tilts by dividing with a normalized linear fit by matching the continuum intensity levels to that of the reference profile (\iref{}).
For the \CaIR{} data, we used spectra synthesized using the RH code \citep{2001ApJ...557..389U} with FAL-C \citep{1985cdm..proc...67A, 1993ApJ...406..319F} model atmosphere at $\mu=0.8$ as \iref{}.
We ensured that there is a good match of synthesized spectra of the \CaIR{} line at $\mu=1$ with BASS 2000 atlas Spectrum, which gave us confidence in using $\mu=0.8$ spectra as \iref{}.
After correcting the continua tilt, we estimated the PSF and straylight fraction.
PSF is assumed to be Gaussian ($\sigma$) for the whole \CaIR{} line spectral range, but straylight fraction ($\nu$) is allowed to vary over the blends of the \SiIIRline{}, the \FeIIRline{} and about the \CaIR{} line core.
To estimate $\sigma$ and $\nu$, we minimised the $\chi^{2}$ distance between the \imeanobs{} and $I_{ref}^{degraded}$ with $\sigma$ and $\nu$.

\begin{equation}
    I_{\mathrm{ref}}^{\mathrm{degraded}}(\lambda) = (1-\nu) I_{\mathrm{ref}}(\lambda) * g(\lambda, \sigma) + \nu * I_{\mathrm{ref}}^{\mathrm{c}} (\lambda)
\end{equation}

where $I_{\mathrm{ref}}^{\mathrm{c}}$ is the value of intensity at the observed predefined far-wing wavelength point.
The absolute SI intensity calibration is done by comparing the intensity of the observed predefined continuum wavelength point with the intensity at degraded \iref{}.

We followed a similar process for estimation of the spectral veil and spectral PSF for the \Halpha{} data, but instead of using synthesized spectrum as \iref{}, we inferred \iref{} using BASS 2000 atlas \Halpha{} spectrum \citep[][]{2009ASPC..405..397P} and center-to-limb variation calculated from $\mu=1$ to $\mu=0.8$ using the RH code with FAL-C model atmosphere.
We used a 6-level Hydrogen atom without a fine structure to synthesize the intensity of \Halpha{} line with the blend of the \FeIVISline{} line.
The atomic parameters of the \FeIVISline{} line are retrieved from Kurucz's line lists \citep{2011CaJPh..89..417K} and synthesized in LTE approximation.

\section{JK Coupling}
\label{sect:jk}

In $J_{1}$-$l$ ($JK$) coupling scheme, a 'parent' level of orbital angular momentum $L_{1}$ and spin $S_{1}$ couples its total angular momentum $J_{1}$ with the orbital angular momentum $l$ of a further electron, to give an angular momentum $K$ which in turn couples with electron's spin to give total angular momentum $J$ \citep[for more details see page 77 of][]{2004ASSL..307.....L}.

The Landé factor for level in $J_{1}$-$l$ ($JK$) coupling is

\begin{equation}
    g_{J_{1}-l} = 1 + \gamma(J, 1/2, K) + \gamma(J, K, 1/2) \gamma(K, J_{1}, l)\gamma(J_{1}, S_{1}, L_{1})    
\end{equation}

where 

\begin{equation}
    \gamma (A, B, C) = \frac{A (A + 1) + B (B + 1) - C (C + 1)}{2 A (A + 1)}
\end{equation}

\section{Quality of fits}
\label{sect:quality}

\begin{figure*}[htbp]
    \centering
    \includegraphics{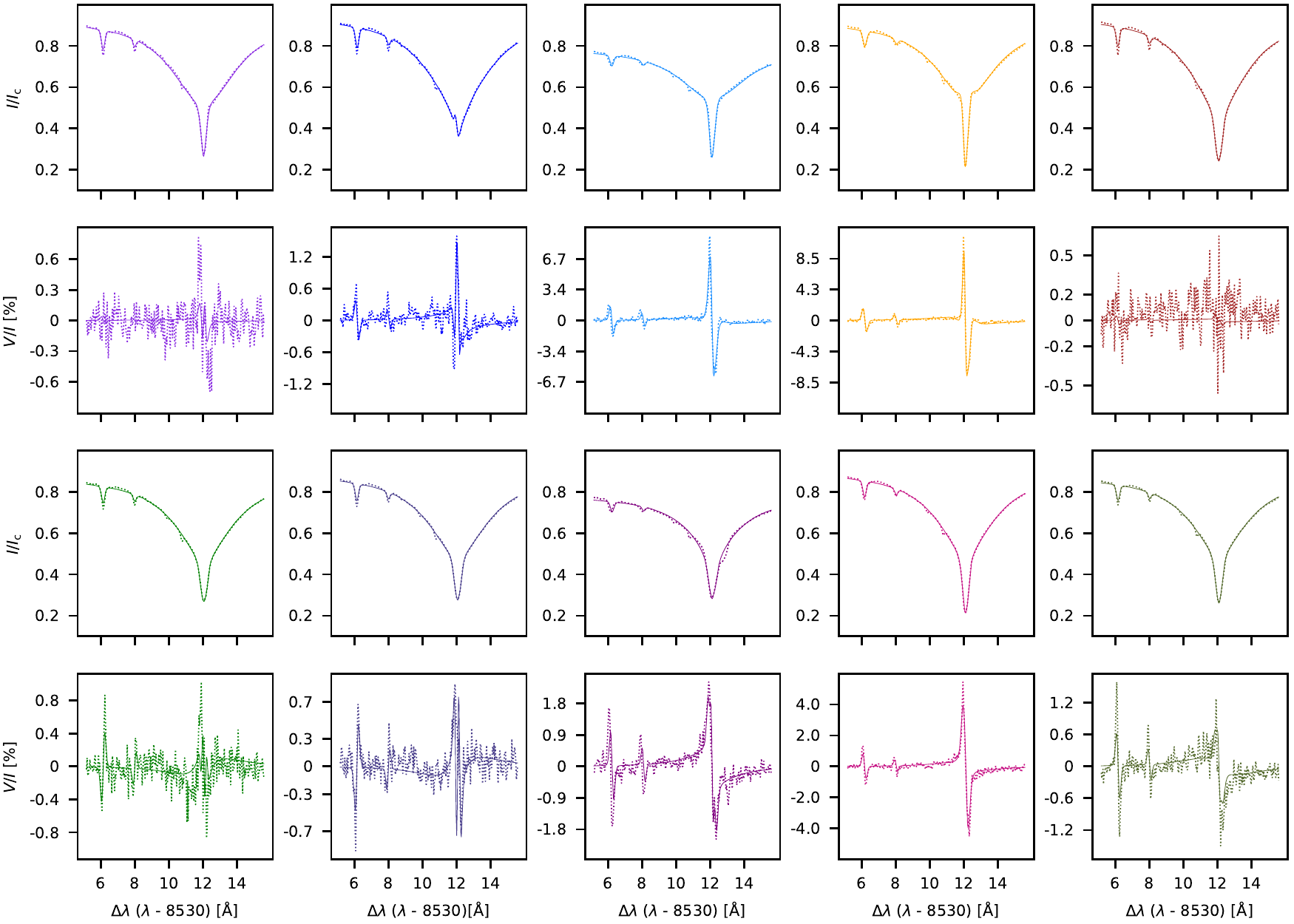}
    \caption{Examples of the observed (dotted lines) and synthesized (solid lines) \CaIR{} Stokes~$I$ and $V$ profiles for the pixels marked by 'x' in Fig.\ref{fig:FOV}.}
    \label{fig:qualityoffits}
\end{figure*}

In Fig.~\ref{fig:qualityoffits} we discuss the match between the observed and synthesized Stokes~$I$ and $V$ for the profiles discussed in the paper.

In general, the synthesized Stokes~$I$ and $V$ profiles show a good match with the observed profiles.
The emission in the blue wing of the \CaIR{} line and reversal in the Stokes~$V$ is very well reproduced in the synthesized profiles (see blue-colored profile).
When the Stokes~$V$ signal is higher than 1\%, a good match is seen in the \SiIIRline{}, \FeIIRline{} and \CaIR{} Stokes~$I$ and $V$ profiles, for example, cyan, maroon, pink and khaki colored profiles.
The opposite sign is very well reproduced in the Stokes~$V$ signal of the \SiIIRline{} and \FeIIRline{} lines and reversal in the \CaIR{} line far wing Stokes~$V$ signal, for example, green and purple colored profile.
When the signal in Stokes~$V$ is less than 0.5\%, the match of the synthesized Stokes~$V$ profile is relatively poor compared to the above cases, for example, purple and brown colored profiles.

\end{document}